# Highly crystalline superconducting $MgB_2$ nanowires formed by electrodeposition

Hideki Abe[*,1,2], Akihisa Miyazoe[2], Keiji Kurashima[1], Kiyomi Nakajima[1], Takeshi Aoyagi[1], Kenji Nishida[1], Noriyuki Hirota[1,2], Takashi Kimura[1], Yoshimasa Sugimoto[2], Tsutomu Ando[2], and Hitoshi Wada[2]

[1]*National Institute for Materials Science (NIMS), Tsukuba, Ibaraki 305-0047, Japan*

[2]*Department of Advanced Materials Science, Graduate School of Frontier Sciences, University of Tokyo, Kashiwanoha 5-1-5, Kashiwa 277-8561, Japan*

Corresponding author:

Hideki Abe

National Institute for Materials Science (NIMS)

1-2-1 Sengen, Tsukuba, Ibaraki 305-0047, Japan

Tel: +81-29-859-2732

Fax: +81-29-859-2301

E-mail: ABE.Hideki@nims.go.jp

Superconducting nanowires can be synthesised in high-throughput chemical routes and hold great promise as a low-dissipative material for superconducting devices[1-9]. The applicability of superconducting nanowires, however, has been limited due to the lack of an adequate method to fit the nanowires into given electronic circuits. One of the biggest obstacles is to connect metal terminals to superconducting nanowires in order to establish electric contacts. One attractive method to surmount this difficulty is to synthesise superconducting nanowires directly upon metal terminals such that the nanowires electrically contact the metal surface. Here we demonstrate that this can be achieved by electrodeposition in molten salts[10-15]. The 39-K superconductor magnesium diboride ($MgB_2$)[16] is electrodeposited to metal surfaces in the form of highly crystalline nanowires. The $MgB_2$ nanowires achieve extensive electric contacts with the metal surfaces. The $MgB_2$ nanowires carry high densities of supercurrents that are comparable to bulk materials. This approach to the electrodeposition of $MgB_2$ nanowires provides a solution to one of the major challenges preventing practical applications of superconducting nanowires.

A variety of superconducting nanowires or nanostripes have been produced by different approaches aimed at obtaining superconducting devices[17-20]. Meander circuits of niobium nitride (NbN) nanostripes, fabricated by ion milling to thin NbN films, have demonstrated potential as a single-photon detector[17-19]. A pair of Mo-Ge nanowires, synthesised by sputtering to DNA templates, has operated as a nanometer-sized superconducting quantum interference device (SQUID)[20]. The Nb- and/or Mo-based devices must operate near the temperature of liquid helium (4.2 K) due to the low superconducting transition temperatures ($T_c$) of these materials. This is not favourable for a practical device, since heavy cryocoolers are required to ensure the operating temperature. Superconducting devices using nanowires or nanostripes of the 39-K superconductor $MgB_2$ may operate at the liquid hydrogen temperature (20 K), significantly reducing cooling costs. The fact that the electron-phonon relaxation time is shorter for $MgB_2$ than for Nb-based superconductors is also an advantage, as it could lead to a higher running speed for $MgB_2$-based superconducting devices[21].

Highly crystalline $MgB_2$ nanowires, preferably with lengths greater than 1 μm, are desirable for the superconducting devices based on electric transport. However, previous attempts to chemically synthesise $MgB_2$ nanowires have resulted in either poor crystallinity or short wires[7-9]. $MgB_2$ nanowires synthesised by gaseous reactions of precursors over 800 °C are longer than 10 μm, but are polycrystalline[7,8]. Conversely, the highly crystalline $MgB_2$ nanowires synthesised by the pyrolysis of bulk $Mg_{1.5}B_2$ at 900 °C are shorter than 200 nm[9]. The low quality of the current $MgB_2$ nanowires is ascribed to the rapid, uncontrollable growth at high temperatures and a high Mg vapour pressure[22]. To obtain $MgB_2$ nanowires under moderate and controllable conditions, we have developed a molten-salt electrodeposition (MSE) technique[10-15]. Unlike the thermal syntheses currently used, MSE is based on the electrolysis of a non-aqueous electrolyte containing the constituent cations of $MgB_2$, $Mg^{2+}$ and $B^{3+}$. Electrolytic co-reduction of $Mg^{2+}$ and $B^{3+}$ leads to the formation of long, highly crystalline $MgB_2$ nanowires at low temperatures and an ambient pressure.

An electrolyte consisting of magnesium chloride ($MgCl_2$) and magnesium borate ($MgB_2O_4$), together with flux materials of sodium chloride (NaCl) and potassium chloride (KCl), was molten at 600 °C under a flow of dry Ar gas. The mole content of the electrolyte was 10:0.1-0.2:5:5 for $MgCl_2$, $MgB_2O_4$, NaCl and KCl, respectively. The chemicals were thoroughly dried and treated under a dry Ar atmosphere. Graphite and platinum (Pt) rods with diameters of 1 mm were used as the anode and reference electrodes, respectively. Either a pure iron or a stainless steel substrate with a thickness of 0.5 mm and a width of 10 mm was used as the cathode. The electrolyte was electrolysed over 10-20 minutes at a constant potential of -1.60 V referenced to the Pt electrode. The electrolysis was terminated by removing the metal substrate from the electrolyte. The metal substrate was washed by sonication in dry methanol to remove the residual electrolyte. The metal substrate was covered with a black electrodeposition film.

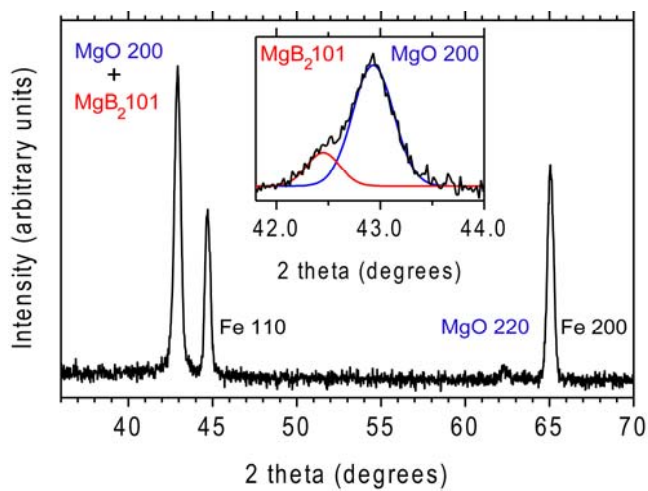

**Figure 1. X-ray diffraction profiles for an electrodeposition film on a metal substrate.** The main panel shows a 2 theta theta scan profile for an electrodeposition film on an iron substrate. The inset shows a close-up of the main panel near 2 theta = 43.0 degrees. The red and blue profiles are simulations for the $MgB_2$ 101 and MgO 200 reflections, respectively.

Figure 1 shows X-ray diffraction profiles for an electrodeposition film on an iron substrate. Besides the reflections from the substrate material, two peaks are recognized at 43.0 and 62.5 degrees. The peak at 62.5 degrees is assigned to the 220 reflection of MgO. The peak at 43.0 degrees consists of the MgO 200 (blue profile) and $MgB_2$ 101 (red profile) reflections, as indicated in the inset. The intensity of the $MgB_2$ 101 reflection is 20 % of the intensity of the MgO 200 reflection. The $MgB_2$ phase in the electrodeposition film is minor in comparison to the MgO phase.

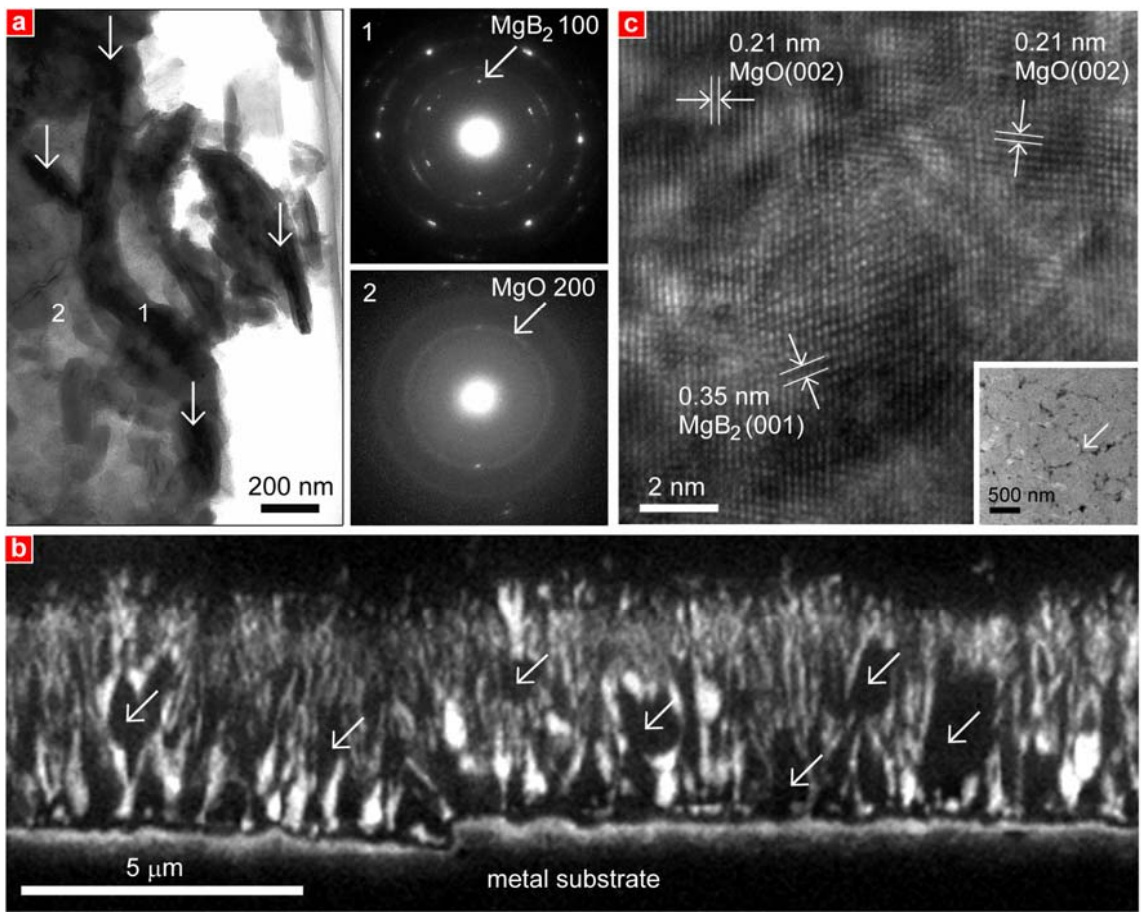

**Figure 2. Nanostructures of electrodeposited films observed with a transmission electron microscope (TEM) and an electron-probe microanalyser (EPMA). a**, A TEM image of an electrodeposited film removed from the substrate. Nanowires with thicknesses of 10-100 nm and lengths of 0.2-1.2 μm are denoted by arrows. The insets show selected area electron diffraction (SAED) patterns corresponding to the areas denoted by 1 and 2. **b**, EPMA compositional mapping of the cross section of an electrodeposited film, visualized using the specific X-ray of boron. The bright area corresponds to a boron-rich phase. The dark areas denoted by arrows are a pure MgO phase. **c**, The main panel shows a high-resolution TEM image of the specimen described in the inset. The inset shows an annular dark field (ADF) image of an in-

plane section of an electrodeposited film. The arrow shows the point at which high-resolution TEM observation was performed.

Figure 2a shows a transmission electron microscope (TEM) image of an electrodeposited film that was removed from an iron substrate. Nanowires with thicknesses of 10-100 nm and lengths of 0.2-1.2 μm appear as dark stripes surrounded by a formless material with a higher brightness. In the centre of the image, there is a nanowire with 100 nm thickness and a length of 1.2 μm. Selected-area electron diffraction (SAED) patterns for this nanowire and the surrounding material are shown in the insets. The sharp, six-fold SAED pattern for the nanowire (inset 1) is assigned to the 001 zone reflection of highly crystalline $MgB_2$. The $MgB_2$ (001) plane is parallel to the long axis of the nanowire. SAED for the surrounding material (inset 2) shows a ring diffraction pattern that is assigned to polycrystalline MgO. Combining the results of TEM and SAED, we conclude that the electrodeposition films consist of highly crystalline $MgB_2$ nanowires and polycrystalline MgO.

Figure 2b shows a cross-sectional image of an electrodeposited film on a stainless steel substrate obtained with an electron-probe microanalyser (EPMA). The average thickness of the electrodeposited film is 4 μm. The $MgB_2$ nanowires are observed as a very fine, columnar texture of a boron–rich phase (bright area) that propagates through the electrodeposited film, almost perpendicular to the substrate surface. An in-plane section parallel to the substrate surface was cut out of an electrodeposited film on an iron substrate using a focused ion beam (FIB). The inset in Figure 2c shows an annular dark field (ADF) image of the in-plane section, which consists of submicron grains bordered by grain boundaries with a darker contrast. SAED showed that the $MgB_2$ phase localizes at the grain boundaries, whereas the submicron grains are a pure MgO phase. The main panel in Figure 2c shows a high-resolution TEM image of the same specimen described in the inset obtained by focusing on one of the grain boundaries. Atomic fringes are observed throughout the image with a wavelength of 0.21 nm, which are assigned to the MgO (200) (*d*-value = 0.2100 nm) and/or $MgB_2$ (101) planes (*d*-value = 0.2126 nm). The centre of the image contains an atomic fringe with a longer wavelength of 0.35 nm. The atomic fringe is assigned to the $MgB_2$ (001) plane (*d*-value = 0.3520 nm). The $MgB_2$ (001) fringe is observed within a restricted area of 4×4 $nm^2$. This area is identified as the cross section of an $MgB_2$ nanowire that grows perpendicular to the substrate surface, with the $MgB_2$ (001) plane aligned parallel to the growth direction. The $MgB_2$ nanowire is embedded in a matrix of MgO.

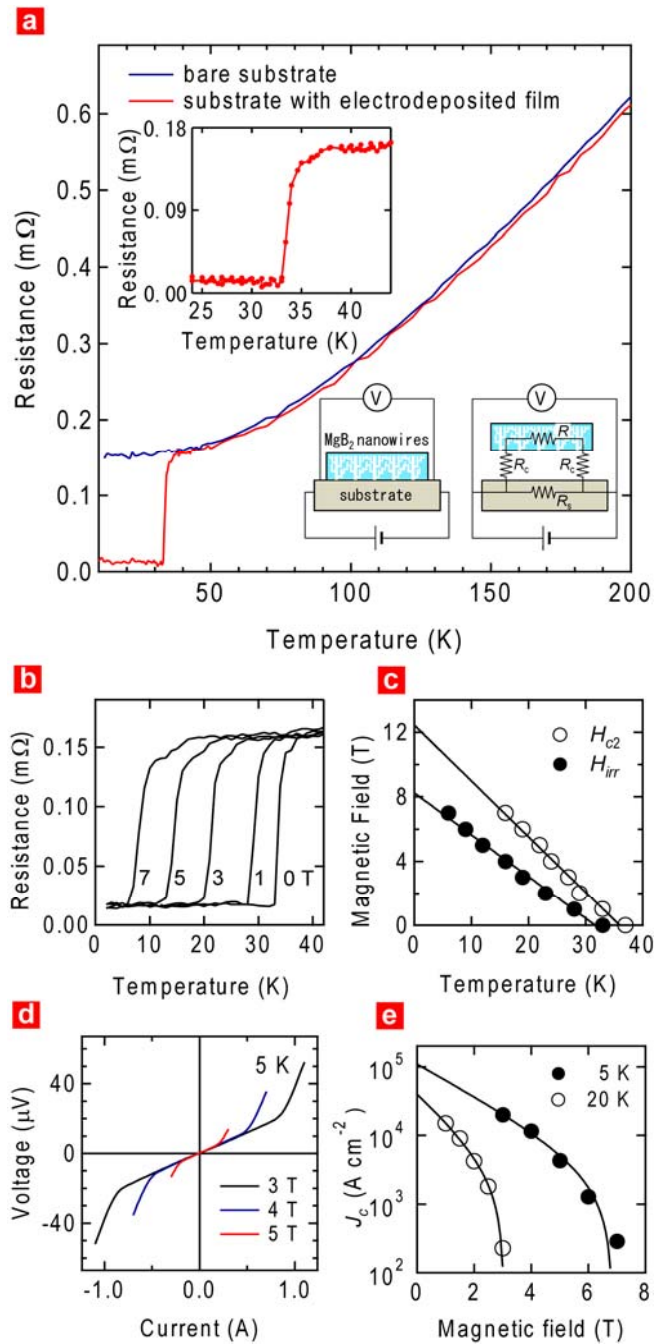

**Figure 3. Electric transport properties of metal substrates with and without electrodeposited films. a**, The main panel shows the resistance/temperature profiles for an iron substrate with an electrodeposited film (red) and for a bare iron substrate (blue). The upper inset shows the resistance/temperature profile near the $T_c$ for the iron

substrate with an electrodeposited film. The lower left and right insets show a schematic picture of the configuration of terminals for transport measurements and an equivalent circuit for the substrate with an electrodeposited film, respectively. **b**, Resistance/temperature profiles for the same substrate described in **a** with an electrodeposited film under finite magnetic fields. **c**, Upper critical field, $H_{c2}$, and irreversibility field, $H_{irr}$, plotted as functions of temperature. **d**, Potential/current profiles under finite magnetic fields. **e**, Effective critical current densities at 5 and 20 K as functions of the magnetic field.

An iron substrate with an electrodeposited film with a thickness of 2.5 μm was cut and shaped into a stripe with a dimension of 1.5×10 mm$^2$ to allow for transport measurements. A bare iron substrate with the same dimension was used as a reference. Both the current and potential terminals were connected to the substrates along the long side. The gap between the potential terminals was 4.5 mm. Figure 3a shows the results of electronic transport measurements on the iron substrates with and without electrodeposited films. The resistance of the bare iron substrate monotonically decreases with a decrease in temperature from 200 to 5 K (blue profile). The resistance of the iron substrate with an electrodeposited film (red profile) traces the blue profile for the bare iron substrate from 200 K down to 40 K. The upper inset shows the resistance/temperature profile below 40 K for the substrate with an electrodeposited film. The resistance drops slightly at an onset temperature of 37.0 K and reaches a minimum at $T_c$ = 33.0 K via an inflection point at 36.2 K. The steep drop in the resistance near the $T_c$ is attributed to the superconducting transition of $MgB_2$ nanowires inside the electrodeposited film.

The finite resistance below the $T_c$ is ascribed to the interfacial resistance between the substrate and film, since the resistivity of electrodeposition films is zero below the $T_c$[13]. The lower insets show the terminal configuration for transport measurements on the substrate with an electrodeposited film, as well as a corresponding equivalent circuit. The resistances of the substrate, electrodeposited film, and the interfacial resistance between the substrate and film are denoted by $R_s$, $R$, and $R_c$, respectively. $R$ is zero below the $T_c$, and $R_s$ and $R_c$ are evaluated at 30 K as 150 and 13 µΩ, respectively. The low $R_c$ shows that the $MgB_2$ nanowires inside the insulating MgO matrix achieve good electrical contact with the metal substrate.

Figure 3b shows the resistance/temperature profiles for the same specimen with an electrodeposited film described in Figure 3a under finite magnetic fields. The magnetic fields were perpendicular to the substrate surface. Both the onset temperature and the $T_c$ decrease with increasing magnetic field. The upper critical field ($H_{c2}$) and the irreversibility field ($H_{irr}$), determined from the onset temperature and $T_c$, respectively, are plotted in Figure 3c as functions of temperature. The upper critical field and the irreversibility field at 0.0 K, $H_{c2}(0)$ and $H_{irr}(0)$, are evaluated as $H_{c2}(0) = 12$ T and $H_{irr}(0) = 8.2$ T by linear extrapolations. The observed $H_{c2}(0)$ lies between the reported values for $MgB_2$ single crystals, $H_{c2}(0) = 25.5$ T ($H$ // (001) plane) and $H_{c2}(0) = 9.2$ T ($H \perp$ (001) plane)[23]. The observed $H_{irr}(0)$ belongs among the lowest literature values, 8-25 T[24], which may show that these $MgB_2$ nanowires are free from impurities or defects that can enhance $H_{irr}(0)$.

Figure 3d shows a current/potential profile for the same specimen described in Figure 3b, again under finite magnetic fields. When the magnetic field is 3 T, the amplitude of potential linearly increases with an increase in the amplitude of the current from 0.0 to 0.8 A. The ohmic dependence of the potential on the current over the range between ± 0.8 A is ascribed to the interfacial resistance between the electrodeposited film and the substrate, $R_c$. The amplitude of the potential grows sharply at a critical current of 0.83 A, which is attributed to the breakdown of the superconducting state of $MgB_2$. The critical current decreases steeply with increasing magnetic field.

On the basis of the equivalent circuit presented in the lower right inset of Figure 3a, the effective critical current that flows through the electrodeposited film is calculated as 90 % of the observed critical current. The effective critical current density for the electrodeposited film, $J_c$, is calculated by dividing the effective critical current by the cross section of the electrodeposited film, 2.5 μm×1.5 mm. Figure 3e shows the plots of the $J_c$ at 5 and 20 K as functions of magnetic field. By extrapolating the plots down to 0.0 T, the $J_c$ at an ambient field is evaluated as $J_c(0) = 1.1 \times 10^5$ and $3.9 \times 10^4$ A cm$^{-2}$ at 5 and 20 K, respectively. These values are comparable to the reported critical current densities for bulk $MgB_2$[24]. Critical current densities for an individual $MgB_2$ nanowire could be much higher than the bulk values, taking into account the low volume fraction of $MgB_2$ nanowires in the insulating MgO matrix.

The results presented in this work demonstrate that MSE is a workable method to synthesise highly crystalline $MgB_2$ nanowires directly on metal surfaces. Outstanding issues and potential limitations to be addressed include the interface between $MgB_2$ nanowires and metal surfaces, and the applicability of MSE to metal terminals of a desired circuit. The chemical and/or thermal stability of the $MgB_2$ nanowires should be also tested to determine whether they are suitable for practical applications. These issues are highly challenging, yet worth the effort, since the low interfacial resistance and the high critical current densities for these $MgB_2$ nanowires will enable the development of the novel electronics of the future.

## Acknowledgments

The authors would like to thank Dr. Kumakura of NIMS for fruitful discussion on the superconducting properties of $MgB_2$ nanowires. This work was supported in part by the Ministry of Education, Culture, Sports, Science and Technology (MEXT) and the Japan Society for the Promotion of Science (JSPS) through Grant-in-Aid 19560845.

## Competing financial interests

The authors declare that they have no competing financial interests.

References

1. Yi, G. & Schwarzacher W. Single crystal superconductor nanowires by electrodeposition. *Appl. Phys. Lett.* **74,** 1746-1748 (1999).

2. Vodolazov, D. Y., Peeters, F. M., Piraux, L., Mátéfi-Tempfli, S. & Michotte, S. Current-voltage characteristics of quasi-one-dimensional superconductors: an s-shaped curve in the constant voltage regime. *Phys. Rev. Lett.* **91,** 157001 (2003).

3. Tian, M. L. *et al.* Synthesis and characterization of superconducting single-crystal Sn nanowires. *Appl. Phys. Lett.* **83,** 1620-1622 (2003).

4. Wang, Y. L., Jiang, X, C., Herricks, T. & Xia, Y. N. Single crystalline nanowires of lead: large-scale synthesis, mechanic studies, and transport measurements. *J. Phys. Chem. B* **108,** 8631-8640 (2004).

5. Xiao, Z. L. *et al*. Tuning the architecture of mesostructures by electrodeposition. *J. Am. Chem. Soc.* **126,** 2316-2317 (2004).

6. Patel, U. *et al*. Synthesis and superconducting properties of niobium nitride nanowires and nanoribbons. *Appl. Phys. Lett.* **91,** 162508 (2007).

7. Wu, Y. Y., Messer, B. & Yang, P. D. Superconducting $MgB_2$ nanowires. *Adv. Mater.* **13,** 1487-1489 (2001).

8. Nath, M. & Parkinson, B. A. Superconducting $MgB_2$ nanohelices grown on various substrates. *J. Am. Chem. Soc.* **129,** 11302-11303 (2007).

9. Ma, R., Bando, Y., Mori, T. & Golberg, D. Direct pyrolysis method for superconducting crystalline $MgB_2$ nanowires. *Chem. Mater.* **15,** 3194-3197 (2003).

10. Abe, H., Yoshii, K. Electrochemical synthesis of superconductive boride $MgB_2$ from molten salts. *Jpn. J. Appl. Phys.* **41,** L685-L687 (2002).

11. Yoshii, K. & Abe, H. Electrical transport properties of bulk $MgB_2$ materials synthesized by electrolysis on fused mixtures of $MgCl_2$, NaCl, KCl and $MgB_2O_4$. *Supercond. Sci. Technol.* **15,** L25-L27 (2002).

12. Yoshii, K. & Abe, H. Electrochemical synthesis of superconductive $MgB_2$ from molten salts. *Physica C* **388-389,** 113-114 (2003).

13. Abe, H., Nishida, K., Imai, M., Kitazawa, H. & Yoshii, K. Superconducting properties of $MgB_2$ films electroplated to stainless steel substrates. *Appl. Phys. Lett.* **85,** 6197-6199 (2004).

14. Abe, H., Yoshii, K., Nishida, K., Imai, M. & Kitazawa, H. Electroplating of the superconductive boride $MgB_2$ from molten salts. *J. Phys. Chem. Solid.* **66,** 406-409 (2005).

15. Miyazoe, A. *et al.* A new approach to $MgB_2$ superconducting magnet fabrication. *J. Phys.* **97,** 012272 (2008).

16. Nagamatsu, J., Nakagawa, N., Muranaka, T., Zenitani, Y. & Akimitsu, J. Superconductivity at 39 K in magnesium diboride. *Nature* **410,** 63-64 (2001).

17. Gol'tsman, G. N. *et al.* Picosecond superconducting single-photon optical detector. *Appl. Phys. Lett.* **79,** 705-707 (2001).

18. Verevkin, A. *et al.* Detection efficiency of large-active-area NbN single-photon superconducting detectors in the ultraviolet to near-infrared range. *Appl. Phys. Lett.* **80,** 4687-4689 (2002).

19. Miki, S. *et al.* Large sensitive-area NbN nanowire superconducting single-photon detectors fabricated on single-crystal MgO substrates. *Appl. Phys. Lett.* **92,** 061116 (2008).

20. Hopkins, D. S., Pekker, D., Goldbart, P. M. & Bezryadin, A. Quantum interference device made by DNA templating of superconducting nanowires. *Science* **308,** 1762-1765 (2005).

21. Khafizov, M., Li, X., Cui, Y., Xi, X. X. & Sobolewski, R. Mechanism of light detection in current-biased superconducting $MgB_2$ microbridges. *IEEE Trans. Appl. Supercond.* **17,** 2867-2870 (2007).

22. Liu, Z. K., Schlom, D. G., Li, Q. & Xi, X. X. Thermodynamics of the Mg-B system: Implications for the deposition of $MgB_2$ thin films. *Appl. Phys. Lett.* **78,** 3678-3680 (2001).

23. Xu, M., *et al.* Anisotropy of superconductivity from $MgB_2$ single crystals. *Appl. Phys. Lett.* **79,** 2779-2781 (2001).

24. Zhou, S. *et al.* Effects of sintering atmosphere on the superconductivity of $MgB_2$. *Supercond. Sci. Technol.* **22,** 045018 (2009), and references therein.